\documentclass[pdflatex,sn-mathphys-num]{sn-jnl}

\usepackage{graphicx}%
\usepackage{multirow}%
\usepackage{amsmath,amssymb,amsfonts}%
\usepackage{amsthm}%
\usepackage{mathrsfs}%
\usepackage{textcomp}%
\usepackage{booktabs}%

\newtheorem{theorem}{Theorem}
\newtheorem{proposition}[theorem]{Proposition}%
\newtheorem{satz}{Satz}
\newtheorem{lemma}{Lemma}

\usepackage[hyphenbreaks]{breakurl}

\usepackage{epsfig}

\def\bra#1{\langle#1|}

\def\ket#1{|#1\rangle}
\def\legendre#1#2{\genfrac{(}{)}{}{}{#1}{#2}}
\def\Q{\mathbb{Q}}

\begin{document}

\title[A Conjecture on Almost Flat SIC-POVMs]{A Conjecture on Almost Flat SIC-POVMs}

\author[1]{\fnm{Ingemar} \sur{Bengtsson}}\email{ibeng@fysik.su.se}
\author*[2]{\fnm{Markus} \sur{Grassl}}\email{markus.grassl@ug.edu.pl}

\affil[1]{
  \orgname{Stockholms Universitet},
  \orgaddress{\street{AlbaNova, Fysikum},\penalty-5000
    \postcode{S-106 91}
    \city{Stockholm},
    \country{Sweden}}}

\affil*[2]{
  \orgdiv{International Centre for Theory of Quantum Technologies},\penalty-5000
  \orgname{University of Gdansk},
  \orgaddress{\street{Jana Bażyńskiego 1a},
    \postcode{80-309}
    \city{Gdansk},
    \country{Poland}}}

\abstract{A well supported conjecture states that SIC-POVMs---maximal
  sets of complex equiangular lines---with anti-unitary symmetry give
  rise to an identity expressing some of its overlaps as squares of
  the (rescaled) components of a suitably chosen fiducial vector. In
  number theoretical terms the identity essentially expresses Stark
  units as sums of products of pairs of square roots of Stark
  units. We investigate whether the identity is enough to determine
  these Stark units. The answer is no, but the failure might be quite
  mild.}




\maketitle

\section{Introduction}

The objects we are interested in are known under various names:
maximal quantum designs, maximal equiangular tight frames, or
Symmetric Informationally Complete POVMs (SIC-POVMs). We call them
SICs. We will assume that they are orbits under a discrete
Weyl--Heisenberg group. The topic of this paper is a simple looking
equation which is assumed to hold for SICs in an infinite series of
special dimensions.

In a complex Hilbert space of dimension $d$ the Weyl--Heisenberg
group is generated by two operators that, in the standard
representation, are given by
\begin{equation}
  Z\ket{r} = \omega^r \ket{r}
  \quad\text{and}\quad
  X\ket{r} = \ket{r+1},
  \qquad\text{where $\omega=e^{\frac{2\pi i}{d}}$.} 
\label{XZ} \end{equation}
The basis vectors $\ket{r}$ are indexed by integers modulo $d$, and it
is easily checked that the operators obey $X^d = Z^d = {\bf 1}$ and
$ZX = \omega XZ$. It is convenient to also define the $d^2$
displacement operators
\begin{equation}
  D_{j,k} = \tau^{jk}X^j Z^k, \qquad
  \text{where $\tau = - e^{\frac{\pi i}{d}}$ and $0\le j,k\le d-1$.}
\end{equation}
They form a unitary operator basis, and it is this property that 
gives the group much of its power. 

Projectively, an orbit under the Weyl--Heisenberg group is a
collection of $d^2$ unit vectors $\ket{\Psi_{j,k}} =
D_{j,k}\ket{\Psi}$, where $\ket{\Psi}$ is known as the fiducial vector
for the orbit. The orbit forms a SIC if and only if there exist phase
factors $e^{i\theta_{j,k}}$ such that
\begin{equation}
  \bra{\Psi}D_{j,k}\ket{\Psi} = \frac{e^{i\theta_{j,k}}} {\sqrt{d+1}}
\qquad\text{for $(j,k)\ne(0,0)$.} \label{eq:def}
\end{equation}
The phase factors $e^{i\theta_{j,k}}$ are left unspecified by the
definition.

There are reasons in quantum engineering and in quantum foundations
why a SIC is a desirable object to have. It came as a surprise to find
that the existence of SICs is connected to a major unsolved problem in
number theory \cite{AFMY}. For our present purposes it is enough to
know that there is a precise conjecture concerning the number field to
which a SIC belongs, and even a conjecture concerning the very numbers
needed to construct a SIC \cite{AFMY,Kopp}.  These numbers are known
as Stark units \cite{StarkIII}. Their importance in number theory is
part of the motivation for this paper, but since we will not rely on
any of their (partly conjectural) properties, we do not present them
here.

SICs are conjectured to exist for all dimensions \cite{Zauner,
  Zauner_english}, and it has been confirmed for all dimensions up
to $196$.  Solutions for $d\le 193$ were found using Andrew Scott's
numerical method \cite{Scott_extending}, dimension $195$ has a
relatively simple exact solution \cite{ApBe2019}, and for $194$ and
$196$ more sophisticated techniques, starting with data from
\cite{AFK}, were used. Exact solutions are known for all dimensions $d
\leq 57$, but existence has not been proven in general.

SICs that are invariant under an anti-unitary symmetry exist in dimensions 
of the form $d = n^2+3$. This was first observed in
the numerical searches \cite{Scott}, but is now well understood
provided one is willing to accept the conjectures about the underlying
number fields. We will focus on this case. For the remainder of 
this introduction we will impose the additional restriction
that the dimension is odd, because it simplifies the main thread of the 
argument. This restriction will be lifted later on.  

A group theoretical argument establishes that in these dimensions any SIC 
that is invariant under an anti-unitary symmetry is unitarily equivalent to a SIC 
having a fiducial vector of the form \cite{ABGHM, BGM}  
\begin{equation}
  \ket{\Psi} = N\ket{{\bf v}},\qquad
  \ket{{\bf v}} = (\sqrt{x_0},v_1, \dots , v_{d-1})^{\rm T}, \label{eq:Ansatz1}
\end{equation}
where 
\begin{equation}
  x_0 = - 2 - \sqrt{d+1}, \qquad
  |v_j|^2 = 1 ,\;
  v_{-j} = - v_j^* ,\qquad 
  N^2 = \frac{1}{d-1-x_0}.\label{Ansatz2}
\end{equation} 
This \emph{almost flat} fiducial vector can be obtained by a discrete
Fourier transformation from a real fiducial vector. Close to one
hundred exact solutions for SICs with anti-unitary symmetry obeying
this Ansatz, or its variant in even dimensions, are known, also in
dimensions much higher than those in which numerical searches are
viable. In these solutions the components $v_i$ are in fact square
roots of Stark units in a number field of surprisingly low degree. We
refer to \cite{ABGHM, BGM} for the details.

The Ansatz we have made for the fiducial vector guarantees that for
$k\ne 0$
\begin{equation}
  \sqrt{d+1}\bra{\Psi}Z^k\ket{\Psi} = 1. \label{eq:Zoverlap}
\end{equation}
In each of a large number of exact solutions it has also been observed 
that 
\begin{equation} \sqrt{d+1} 
  \bra{\Psi}X^{-2j}\ket{\Psi} = \frac{\psi_j^2}{|\psi_j|^2}
  \qquad\text{for $j\ne 0$,} \label{eq:Xoverlap0}
\end{equation}
where $\psi_j$ are the components of the unit vector $\ket{\Psi}$. By writing 
out the definitions one finds the equivalent form  
\begin{equation}
  \bra{{\bf v}} X^{-2j}\ket{{\bf v}} = (\sqrt{d+1} + 1)v_j^2
  \qquad\text{for $j\ne 0$.}\label{eq:Xoverlap}
\end{equation} 
We will refer to either of these equations \eqref{eq:Xoverlap0} and
\eqref{eq:Xoverlap} as the \emph{$X$-overlap equation}.  The
conjecture referred to in the title is that this equation holds for
all almost flat SIC fiducial vectors in odd dimension.  Given how the
operator $X$ is represented, it is an equation for the periodic
autocorrelation of the components of the fiducial vector.  The
equation guarantees that these overlaps have the correct absolute
values, but at the same time it is clearly a stronger
statement. Number theorists should notice that in the known solutions
for SICs the phase factors $v_j^2$ are Stark units, which means that
the equation translates into identities connecting certain Stark units
to sums of products of pairs of square roots of Stark units
\cite{ABGHM, BGM}.

We also note that, unlike the SIC conditions, the $X$-overlap equation
is sensitive to multiplying the vector by a phase
factor, i.\,e., a complex number of modulus one.  We have chosen the first 
component $\sqrt{x_0}$ of the vector $\ket{\bf v}$ to be purely imaginary. 
Had it been chosen to be purely real there would be a change of sign 
on the right hand side of \eqref{eq:Xoverlap}.

The question that we are raising here is whether the $X$-overlap 
equation, together with the Ansatz for the fiducial vector, is enough to 
guarantee that the full SIC equations \eqref{eq:def} hold as well? 

A positive answer would turn the conjecture in the title into an
if-and-only-if statement.  We will however answer the question in the
negative.  We will find non-SIC solutions to the $X$-overlap equation
in some dimensions, but they occur in all prime dimensions and we have
an explicit description of them.  Hence it remains possible that the
obstruction may be quite mild.

The paper is organized as follows: In Section
\ref{sec:detailed_Ansatz} we bring the Ansatz to a form more
convenient for calculations, give a sketch of how to lift the
restriction to odd dimensions, and explain why the question we have
asked is reasonable and interesting.  In Section \ref{sec:Groebner} we
report on Gr\"obner basis calculations that have enabled us to find
the full set of solutions to the $X$-overlap equation in a few
dimensions of the form $d = n^2+3$, namely $7$, $12$, $19$, $28$,
$39$, $52$, $67$, and $199$. (In most of these dimensions we did impose 
a permutation symmetry, known to be there for SICs, on the vector.) 
We find a spurious, non-SIC solution when 
$d = 67$ and $d=199$.  In Section \ref{sec:Legendre_vecs} we give similar 
non-SIC solutions to the equation in all dimensions of the form $d = p
\equiv 3 \mod 4$ where $p$ is a prime.
These spurious solutions are so-called Legendre vectors, as defined by
Khatirinejad \cite{Mahdad}.

\section{A closer look at the Ansatz and the $X$-overlap equation}\label{sec:detailed_Ansatz}

A key fact about all known exact solutions for SICs in dimensions
$d>3$ is that the phase factors appearing in the overlaps are
algebraic units. To simplify the discussion a little, let us restrict
ourselves mainly to minimal SICs (where the associated number field
has minimal degree) in the case when $d =n^2 + 3 = p$ where $p$ is an
odd prime.  This prime is necessarily equal to $3 \bmod 4$ and equal
to $1 \bmod 3$, and a long standing conjecture \cite{hardy} implies
that an infinite sequence of such primes exist. In all cases that have
been examined one finds that the `large' overlap phase factors,
i.\,e., those coming from the overlaps $\bra{\Psi}D_{j,k}\ket{\Psi}$
with both $j$ and $k$ being nonzero, are square roots of Stark units
forming a single Galois orbit in a number field whose degree grows
with dimension like $(d-1)^2$. The components $v_j$ of the fiducial
vector on the other hand form a single Galois orbit in a subfield
whose degree grows like $d-1$. In both cases there is a permutation
symmetry of order $3\ell$ (where $\ell$ is a known dimension dependent
integer) acting, so that the length of the Galois orbit of the
components is actually only $(d-1)/3\ell$. For ease of reference we
refer to the field holding the fiducial vector as the `small' ray
class field, and to the field holding all the SIC overlaps as the
`large' ray class field. For full details see \cite{ABGHM,BGM}.

It is possible to build the permutation symmetry into our Ansatz for the 
fiducial vector, but it is also interesting to see whether the $X$-overlap 
equation is enough to force the vector to obey the symmetry. In the 
Gr\"obner basis calculations that we report in Section \ref{sec:Groebner} 
we dealt with the small dimensions $d =7$ and 19 without imposing 
the symmetry, but we did impose it in the remaining examples. 

A slight complication must be mentioned in connection with our Ansatz,
namely that taking the square root of $x_0$ causes the degree of the
number field to increase by a factor of two. This can be avoided by a
rescaling
\begin{equation}
  \ket{\Psi} = N
  \left(\begin{array}{c}
    \sqrt{x_0} \\
    v_1 \\ \vdots \\ v_{d-1}
  \end{array} \right) \rightarrow
  N'\left(\begin{array}{c}
    x_0 \\ 
    \sqrt{x_0}v_1 \\
    \vdots \\ 
    \sqrt{x_0}v_{d-1}
  \end{array}\right).\label{eq:rescaled_vector}
\end{equation}
The components of the rescaled vector do not need the extra extension
of the number field. In the simplest, minimal, case that was discussed
in \cite{ABGHM} the $v_j$ turn out to be square roots of Stark units
in the `small' ray class field.  One needs the rescaling if one wants
to take their square roots within that ray class field.  Something
similar is true in those non-minimal cases that have been examined. In
the calculations that we report in Section \ref{sec:Groebner} the
rescaled version of the Ansatz was used.

Another complication occurs when the dimension is even. If so it must
be true that $d = n^2+3 = 4k$ where $k$ is odd. The Weyl--Heisenberg
group will then split as a direct product and the Hilbert space splits 
correspondingly into a tensor product. It is the dimension four
factor that causes some complications.  The resolution offered in
\cite{BGM} is to use an unusual representation of the group in the
dimension four factor, ensuring that the symmetries of the fiducial
vector are still given by a permutation matrix. The $X$-overlap
equation then takes the form given in eqs. (94)--(95) in \cite[Section
  8]{BGM}.  Since the details were fully explained there, we do not
repeat them here.  Unless stated otherwise we will focus our
discussion on the case when the dimension is odd, even though some
results for even dimensions are given in Section \ref{sec:Groebner}.

These preliminaries attended to, we turn to the $X$-overlap equation \eqref{eq:Xoverlap}. 
The exponent of the operator $X$ on the left hand side looks a little peculiar. 
It can be made more natural by observing that our Ansatz for the fiducial vector 
ensures the following equivalence:

\begin{proposition}\label{prop1}
  Let $\psi_i$ denote the components of a vector $\ket{\Psi}$ that
  obeys the Ansatz given in eqs. \eqref{eq:Ansatz1} and
  \eqref{Ansatz2}. Then for $j\ne 0$
  \begin{equation}
    \sqrt{d+1}
    \bra{\Psi}X^{-2j}\ket{\Psi} = \frac{\psi_j^2}{|\psi_j|^2} 
 \quad\Longleftrightarrow\quad
   \sum_{k=1}^{d-1}\bra{\Psi}D_{j,k}\ket{\Psi} =
 - \sqrt{d+1}\bra{\Psi}X^j\ket{\Psi}.\label{eq:identitet}
  \end{equation}
\end{proposition}

\begin{spiproof}
We use the explicit matrices in the standard representation 
of the Weyl--Heisenberg group, and find 
\begin{alignat}{11}
  \left( \sum_{k=0}^{d-1}D_{j,k}\right)_{r,s}\!\!\!
  &{}= \left( X^j\sum_{k=0}^{d-1}\tau^{jk}Z^k\right)_{r,s}\!\!\!
  &&{}={}&&\sum_{t=0}^{d-1}\delta_{r,t+j}\sum_{k=0}^{d-1}\tau^{k(j+2t)}\delta_{t,s} \nonumber  \\
 &&&{}=d& &\sum_{t=0}^{d-1}\delta_{r,t+j}\delta_{2t,-j}\delta_{t,s}
  =d\delta_{r,2^{-1}j}\delta_{s,-2^{-1}j}.
\end{alignat}
Note the occurrence of the inverse $2^{-1}$ modulo $d$. After substituting 
$-2j$ for $j$ it follows that 
\begin{equation}
  \bra{\Psi}X^{-2j}\ket{\Psi}
    + \bra{\Psi}\sum_{k=1}^{d-1}D_{-2j,k}\ket{\Psi}
    = d\psi_{-j}^*\psi_j = - \frac{d}{d+1 + \sqrt{d+1}}\frac{\psi_j^2}{|\psi_j|^2},
\end{equation}
where the Ansatz for the components of the fiducial vector was used in the last 
step. A minor calculation concludes the proof. 
\end{spiproof}

From the number theoretical point of view we find, assuming that we
are dealing with minimal SICs, that the original $X$-overlap equation
is an identity that is obeyed (at least in all cases examined) by
square roots of Stark units in the `small' ray class field. In its
equivalent form of Proposition \ref{prop1} it connects Stark units in
the `small' ray class field to sums of products of square roots of
Stark units in the `large' ray class field. Proving that these
identities always hold is presumably very hard when using only the
definition of the Stark units.

Finally, our initial hypothesis was that the $X$-overlap equation
imposed on our Ansatz has only SIC vectors as solutions. This
hypothesis will be shot down in Section \ref{sec:Groebner}, but let us
nevertheless explain why it seemed reasonable. The first observation
is that the $X$-overlap equation is clearly stronger than the simple
statement that
\begin{equation}
  |\bra{\Psi}X^j\ket{\Psi} |^2 = \frac{1}{d+1},\quad j=1,\ldots,d-1,   \label{eq:Xnaiv}
\end{equation}
which is what a subset of the equations in \eqref{eq:def} defining a
SIC would say. In fact equations \eqref{eq:Xnaiv} have spurious
non-SIC solutions already in dimension $d = 7$, while the $X$-overlap
equation does not. But an obvious objection is that, because the
operator $X$ is just a cyclic permutation matrix and the components of
the vector belong to the `small' ray class field, the hypothesis
assumes that the SIC conditions can be formulated entirely in that
`small' field, even though the SIC itself lives in a much larger
field. Such an objection does not withstand closer examination. One may use 
the discrete Fourier transform to define and calculate the quantities \cite{Mahdad, ADF} 
\begin{equation}
  G(i,k)
    = \frac{1}{d}\sum_{j=0}^{d-1}\omega^{kj}|\bra{\Psi}X^iZ^j\ket{\Psi} |^2
     = \sum_{r=0}^{d-1}\psi^*_{r+i}\psi^*_{r+k}\psi_r\psi_{r+i+k}.
\end{equation}
The roots of unity drop out of the expression, for any vector with 
components $\psi_r$. But it is also true that 
\begin{equation}
  |\bra{\Psi}X^iZ^j\ket{\Psi} |^ 2
    = \frac{d\delta_{i,0} \delta_{j,0} + 1}{d+1}
    \quad\Longleftrightarrow\quad
    G(i,k) = \frac{\delta_{i,0} + \delta_{k,0}}{d+1}. \label{eq:Gik2}
\end{equation} 
This means precisely that the SIC conditions can be checked by means of 
a calculation involving only the components of the fiducial vector. Moreover 
the $Z$-overlap equation \eqref{eq:Zoverlap} is built into our Ansatz, and 
this is enough to guarantee that all the non-homogeneous equations hold. 

The question then is whether the restriction imposed by the
$X$-overlap equation is enough to force the components of the fiducial
vector to obey the homogeneous equations for the $G(i,k)$? In Section
\ref{sec:Legendre_vecs} we will find that the answer is no. This
result was inspired by investigating a few examples for small
dimensions, as reported in Section \ref{sec:Groebner}.  To some
extent, those examples suggest that the obstruction
 might be rather mild.

\section{Gr\"obner basis calculations}\label{sec:Groebner}

We will describe our results in detail for the three prime dimensions
$d = 7$, $19$, and $67$.  For the remaining cases we will simply state
the key results. As it turns out it is in dimension $d = 67$ that our
hypothesis fails. The labeling of solutions follows \cite{Scott} and
\cite{Scott_extending}. A useful piece of background information is
that with high confidence every SIC, with or without anti-unitary
symmetry, has been identified by numerical searches for $d \leq 90$.

\subsection{Dimension $7$}
We start with $d=7$, the smallest prime dimension of the form $d=n^2+3$. We do not 
impose the permutation symmetry on the Ansatz. 
From \eqref{eq:rescaled_vector}, the non-normalized fiducial vector has
the form
\begin{alignat}{5}
  {\bf x}=(x_0,x_1,\ldots,x_6)^{\rm T}=(x_0,\sqrt{x_0}v_1,\ldots,\sqrt{x_0} v_6)^{\rm T},\label{eq:vector_x}
\end{alignat}
where $x_0=-2-\sqrt{d+1}=-2-2\sqrt{2}$, $|v_j|^2=1$, and $v_{-j}=-v_j^*$ by \eqref{Ansatz2}.
From the condition $v_{-j}=-v_j^*=-1/v_j$ of the Ansatz, we get the
equations
\begin{alignat}{5}
  x_1 x_6 = x_2 x_5 = x_3 x_4 = -x_0.
\end{alignat}
The complex conjugate of the vector ${\bf x}$ has the form
\begin{alignat*}{5}
  {\bf x^*}=(x_0,x_1^*,\ldots,x_6^*)^{\rm T}
    &{}=(x_0,-\sqrt{x_0}/v_1,\ldots,-\sqrt{x_0}/v_6)^{\rm T}\\
    &{}=(x_0,-x_0/x_1,\ldots,-x_0/x_6)^{\rm T}\\
    &{}=(x_0,x_6,x_5,\ldots,x_1)^{\rm T}.
\end{alignat*}
For $x_0$ we have the quadratic equation $(x_0+2)^2=d+1$. Finally, we
add the $X$-overlap equation from \eqref{eq:Xoverlap0}.
A Gr\"obner basis for the ideal generated by these equations with
respect to the lexicographic order with $x_1>x_2>\ldots>x_6>x_0$ is
\begin{alignat}{5}
  \{&  x_1 + 35/64 x_6^{13} x_0 + 169/64 x_6^{13} - 41/16 x_6^6 x_0 - 99/8 x_6^6,\notag\\
  &  x_2 - 157/448 x_6^{12} x_0 - 379/224 x_6^{12} + 95/56 x_6^5 x_0 + 115/14 x_6^5,\notag\\
  &  x_3 - 65/448 x_6^{11} x_0 - 157/224 x_6^{11} + 33/56 x_6^4 x_0 + 81/28 x_6^4,\notag\\
  &  x_4 - 65/448 x_6^{10} x_0 - 157/224 x_6^{10} + 5/7 x_6^3 x_0 + 51/14 x_6^3,\notag\\
  &  x_5 - 19/224 x_6^9 x_0 - 23/56 x_6^9 + 9/28 x_6^2 x_0 + 17/14 x_6^2,\notag\\
  &  x_6^{14} + 4 x_6^7 x_0 - 8 x_6^7 - 10816 x_0 + 8960,\notag\\
  &  x_0^2 + 4 x_0 - 4\}.\label{eq:ideal7}
\end{alignat}
The ideal is zero-dimensional and the variety has $28$ points. Over
the rationals, it splits into two irreducible components with $4$ and
$24$ points, respectively. A Gr\"obner basis for the first component
is
\begin{alignat}{5}
 \{ & x_1 + x_6 + \frac{1}{2} x_0 - 1,  x_2 + x_6 + \frac{1}{2} x_0 - 1, x_4 + x_6 + \frac{1}{2} x_0 - 1,\notag\\
    & x_3 - x_6, x_5 - x_6,\notag\\
    & x_6^2 + \frac{1}{2} x_6 x_0 - x_6 - x_0,
      x_0^2 + 4 x_0 - 4\}.\label{eq:ideal7a}
\end{alignat}
In this case $x_1=x_2=x_4$ and $x_3=x_5=x_6$, i.\,e., the solution has
a permutation symmetry.  We omit the explicit equations for the second
component.

Note that the equation $(x_0+2)^2=d+1$ has two solutions,
corresponding to the choice of the sign of the square root of $d+1$. 
In our Ansatz, we have used the assumption that $\sqrt{d+1}>0$, which
implies that $x_0<0$ and hence $\sqrt{x_0}$ is purely imaginary. This
assumption reduces the number of actual solutions by a factor of two.
The other branch of the square root $\sqrt{d+1}$  yields the `ghost SICs', in the
terminology of \cite{AFK}, for the small component.

Using $\beta=\sqrt{-2\sqrt{2}-1}$, we obtain the SIC fiducial vectors
\begin{alignat*}{5}
{\bf x}_1=\frac{1}{2}\begin{pmatrix}
        -4\sqrt{2} - 4\\
        \sqrt{2}(-\beta + 1) + 2\\
        \sqrt{2}(-\beta + 1) + 2\\
        \sqrt{2}(\beta + 1) + 2\\
        \sqrt{2}(-\beta + 1) + 2\\
        \sqrt{2}(\beta + 1) + 2\\
        \sqrt{2}(\beta + 1) + 2
\end{pmatrix}
\qquad\text{and}\qquad
{\bf x}_2=\frac{1}{2}\begin{pmatrix}
        -4\sqrt{2} - 4\\
        \sqrt{2}(\beta + 1) + 2\\
        \sqrt{2}(\beta + 1) + 2\\
        \sqrt{2}(-\beta + 1) + 2\\
        \sqrt{2}(\beta + 1) + 2\\
        \sqrt{2}(-\beta + 1) + 2\\
        \sqrt{2}(-\beta + 1) + 2
\end{pmatrix},
\end{alignat*}
which differ by the choice of the sign of $\beta$. Up to
normalization, both vectors are SIC fiducial vectors.  It
turns out that the $12$ solutions with $x_0<0$ from the other
irreducible component can be obtained multiplying ${\bf x}_1$ and
${\bf x}_2$ by the operator $Z^k$, $k=1,\ldots,6$; the same applies to
the solutions with $x_0>0$. These solutions had to be there because 
\begin{equation} \sqrt{d+1} 
  \bra{\Psi}Z^{-1}X^{-2j}Z\ket{\Psi} = \omega^{2j}\frac{\psi_j^2}{|\psi_j|^2}.
\end{equation}
Hence the vector $Z\ket{\Psi}$ obeys the identity \eqref{eq:Xoverlap0} whenever 
the vector $\ket{\Psi}$ does. 

\subsection{Dimension $19$}
The next prime dimension of the form $n^2+3$ is $d=19$.
For the vector ${\bf x}=(x_0,\ldots, x_{18})^{\rm T}$, our Ansatz
implies the equations
\begin{alignat}{5}
  x_1 x_{18} = x_2 x_{17} = \ldots = x_9 x_{10} = -x_0  
\end{alignat}
with $(x_0+2)^2=d+1=20$. Together with the $X$-overlap equation
\eqref{eq:Xoverlap0}, once again we obtain a zero-dimensional ideal,
now with $304$ points. Over the rationals, the variety has four
irreducible components with $4$, $12$, $72$, and $216$ points,
respectively. For half of the points, $x_0<0$.

For the smallest component, we have two solutions with $x_0<0$ and
\begin{alignat*}{5}
&&  x_1=x_4=x_5=x_6=x_7=x_9=x_{11}=x_{16}=x_{17}={}& \beta-1\\
\text{and}\quad &&  x_2=x_3=x_8=x_{10}=x_{12}=x_{13}=x_{14}=x_{15}=x_{18}={}&-\beta-1,
\end{alignat*}
where the two solutions differ by the choice of the sign of
$\beta=\sqrt{-2\sqrt{5}-1}$. Those solutions are SIC fiducial vectors
with a cyclic permutation symmetry of order $9$ with $x_{4j}=x_j$. The
solution is unitarily equivalent to the solution labeled $19e$.

For the component with $12$ points, we get six solutions with $x_0<0$ and
\begin{alignat*}{7}
& x_1=x_7=x_{11},
&\quad& x_2=x_3=x_{14},
&\quad& x_4=x_6=x_9,\notag\\
& x_5=x_{16}=x_{17},
&\quad& x_8=x_{12}=x_{18},
&\quad& x_{10}=x_{13}=x_{15}.
\end{alignat*}
This shows a permutation symmetry of order $3$ with $x_{7j}=x_j$.  The
solution is unitarily equivalent to the solution labeled $19d$. It is
a non-minimal SIC with anti-unitary symmetry.

The solutions from the two larger components can again be obtained
applying operators $Z^k$, $k=1,\ldots,18$, to the solutions in the two
small components. Again the symmetry of these solutions is no longer a
permutation, but the vectors are elements of the same SICs as the
solutions from the smaller components.

\subsection{Dimension $67$}
For dimensions $7$ and $19$, our Ansatz only assumed that the fiducial
vector is almost flat, i.\,e., $|v_j|^2=1$, and that complex conjugation
acts as $v_j^*=-v_{-j}$. In both cases, this Ansatz together with the
$X$-overlap equation yields solutions with permutation symmetry, as
well as solutions that are obtained applying powers of the operator
$Z$.

For the next prime dimension of the form $n^2+3$, $d=67$, the
calculations become lengthier and we impose an additional permutation
symmetry of order $3$ related to the order-three symmetry conjecture
by Zauner \cite{Zauner, Zauner_english}. For prime dimensions
$d=p\equiv 1 \bmod 3$, Zauner's symmetry is always conjugate to a
permutation symmetry. For $d=67$, we obtain the additional equations
$x_j= x_{29j}$ for $1\le j<67$. Computing a Gr\"obner basis using
Magma \cite{Magma} over the rationals took about $9$ days. Again we
obtain a zero-dimensional ideal with $92$ points. The variety splits
into three irreducible components with $4$, $44$, and $44$ points,
respectively. The two larger components give rise to $22$ SIC fiducial
vectors each, corresponding to the orbits labeled $67a$ and
$67b$. Here $67a$ is a minimal SIC and $67b$ is non-minimal, even
though they sit in the same number field.

For the smallest component, we get two solutions with $x_0<0$, given
by 
\begin{alignat*}{50}
  x_j=\begin{cases}
   \beta-1,&\text{if $j\ne 0$ is a square modulo $67$;}\\
  -\beta-1,&\text{if $j\ne 0$ is not a square modulo $67$,}
  \end{cases}
\end{alignat*}
where the two solutions differ by the choice of the sign of
$\beta=\sqrt{-2\sqrt{17}-1}$. In this case, the solutions to the
$X$-overlap equation are \emph{not} SIC fiducial vectors. 
Hence our initial hypothesis turns out to be wrong in general. 

\subsection{Further results}
We have also tested the $X$-overlap equation for some composite dimensions
of the form $d=n^2+3$, as well as for dimension $199$.

\subsubsection{Dimension $39$}
The first composite odd dimension of the form $n^2+3$ is $d=39$. As
for dimension $67$, we additionally impose a permutation symmetry of
order $3$. We obtain a zero-dimensional ideal with $96$ points. The
variety splits into two irreducible components with $32$ and $64$
points, respectively. The small component yields $16$ SIC fiducial
vectors related to the minimal solutions labeled $39i$ and $39j$, as
well as $16$ ghost fiducials. The solutions from the larger component
are obtained multiplying the solutions from the small component by the
operators $Z^{13}$ and $Z^{26}$. Note that these operators share the
same permutation symmetry of order $3$ as the solutions, since the
dimension is divisible by $3$.

\subsubsection{Dimension $199$}
For the prime dimension $199=14^2+3$, there is an almost flat SIC
fiducial vector that has a permutation symmetry of order $9$
\cite{ABGHM}, as well as an unpublished almost flat SIC fiducial
vector with a permutation symmetry of order $3$.  Using only the
symmetry of order $3$, we have not been able to solve the $X$-overlap
equation.  However, imposing a permutation symmetry of order $9$, we
are able to compute a Gr\"obner basis for the $X$-overlap equation in
a bit more than $9$ days.  Once more, the ideal has dimension zero.
The variety has two irreducible components with $4$ and $44$ points,
respectively.  The $22$ points with $\sqrt{x_0}<0$ in the larger
component correspond to the solution $199a$, and the other $22$ points
to ghost fiducials. The $4$ solutions in the smaller component are not
SIC fiducials.

\subsubsection{Even dimensions}
As briefly mentioned before, for even dimensions the $X$-overlap
equation splits into two, given in eqs. (94)--(95) in \cite[Section
  8]{BGM}. We also impose a permutation symmetry of order three, which
is part of the Ansatz used in \cite{BGM}.  In all cases, we obtain a
zero-dimensional ideal.

For $d=12=4\times 3$, there are two irreducible components with $8$
and $16$ points, respectively. The solutions correspond to the minimal
SIC labeled $12b$, and the corresponding ghost fiducials.  The larger
variety is obtained multiplying the solutions from the smaller
component with the operators $Z^4$ and $Z^8$. The situation is similar
to dimension $39$, both dimensions are divisible by $3$.

For $d=28=4\times 7$, there is only one irreducible component with
$24$ points. The solutions correspond to the SIC fiducial vector
labeled $28c$, and the corresponding ghost fiducials.

Similarly, for $d=52=4\times 13$, there is only one irreducible component
with $48$ points. The solutions correspond to the orbit labeled
$52d$ and the corresponding ghost fiducials.

\section{Legendre vectors}\label{sec:Legendre_vecs}

Khatirinejad \cite{Mahdad} proposed a stronger version of our Ansatz
for dimensions of the form $d = p\equiv 3 \mod 4$. This covers all the
cases for which $d = n^2 + 3$ equals a prime $p$ (such as $d = 7, 19,
67, \dots$, but also many other dimensions such as $d = 11, 23, 31,
\dots$). We define a \emph{Legendre vector} as a vector whose components obey 
our equations \eqref{eq:Ansatz1} and \eqref{Ansatz2} with the additional 
requirement that
\begin{equation}
  v_j = \begin{cases}
    \sqrt{x_0}, & \text{if $j=0$;}\\
    x_1, &\text{if $j$ is a quadratic residue;}\\
    - 1/x_1, & \text{otherwise.}
\end{cases} \label{eq:Ansatz3}
\end{equation}
Our Ansatz implies that $x_1^* = 1/x_1$ so that $x_1$ is a phase
factor.  Our Ansatz also requires $v_{-j} = - v_j^*$, and the
definition of Legendre vectors is consistent with this because $-1$ is
not a quadratic residue modulo $p$ when $p\equiv 3\bmod 4$. In terms
of the Legendre symbol
\begin{equation}\label{Leg1}
  \legendre{-1}{p}=-1\qquad\text{if $p\equiv 3\bmod 4.$} 
\end{equation}
Recall that the Legendre symbol $\legendre{n}{p}$ is multiplicative.
It equals $+1$ if $n$ is a quadratic residue modulo $p$, and equals
$-1$ if $n$ is not a quadratic residue.

The inspiration for Legendre vectors came from Appleby's exact SIC
solutions in dimensions $7$ and $19$, where SIC fiducials of this form
exist (with symmetries of order $3$ and $9$, respectively)
\cite{Marcus}. They were indeed recovered in Section
\ref{sec:Groebner}.  Khatirinejad proved that the sequence of
compatible prime dimensions for which these Legendre vectors might be
SIC fiducials has zero density in the set of all primes that are equal
to $3$ modulo $4$.  With the benefit of hindsight the number
theoretical conjectures imply that it is only in dimensions $7$ and
$19$ that the symmetry group of the SIC can be large enough to ensure
that only two distinct phase factors occur in the vector.  However, we
will now show that Legendre vectors solving the $X$-overlap equation
exist in all dimensions $d = p \equiv 3\bmod 4$.

To do so we need two theorems by Perron \cite{Perron52}. In his terminology, a
``Rest'' (plural ``Reste'') is either a quadratic residue or equal to
$0$. The other elements are referred to as ``Nichtreste''.
\begin{satz}
  Let $p\equiv 3\bmod 4$ and $a$ be any coprime integer.  Then there
  are $(d+1)/4$ Reste and $(d+1)/4$ Nichtreste among the numbers $r_i
  + a$, where $r_i$ runs over all the $(d+1)/2$ Reste.
\end{satz}
\begin{satz}
  Let $p\equiv 3\mod 4$ and $a$ be any coprime integer.  Then there
  are $(d+1)/4$ Reste and $(d-3)/4$ Nichtreste among the numbers $n_i
  + a$, where $n_i$ runs over all the $(d-1)/2$ Nichtreste.
\end{satz}
\noindent
Satz 2 is a corollary of Satz 1. Perron's proof of Satz 1 is elementary. 

We use Perron's theorems to prove a lemma:
\begin{lemma}\label{lemma1}
  Given a vector $\ket{{\bf v}}$ that obeys the Ansatz
  \eqref{eq:Ansatz1}, \eqref{Ansatz2}, and \eqref{eq:Ansatz3}. If
  $\legendre{j}d = 1$ and $d \equiv 3\bmod 8$, then
\begin{equation}
  \bra{{\bf v}}X^{-2j}\ket{{\bf v}} = \frac{d-3}{2} - \frac{d-3}{4}
  \frac{1}{x_1^2} - \frac{d+1}{4}x_1^2 + \frac{2\sqrt{x_0}}{x_1}.
  \label{fall1}
\end{equation} 
If $\legendre{j}d = 1$ and $d\equiv 7\bmod 8$, then 
\begin{equation}
  \bra{{\bf v}}X^{-2j}\ket{{\bf v}}
  = \frac{d-3}{2} - \frac{d+1}{4}\frac{1}{x_1^2} - \frac{d-3}{4}x_1^2 -  2\sqrt{x_0}x_1.
\label{fall2} \end{equation}
If $\legendre{j}{d}=-1$, we must perform the substitution 
$x_1 \mapsto -1/x_1$ in these equations. 
\end{lemma}

\begin{figure}[hb]
\centerline{ \hbox{
                \epsfig{figure=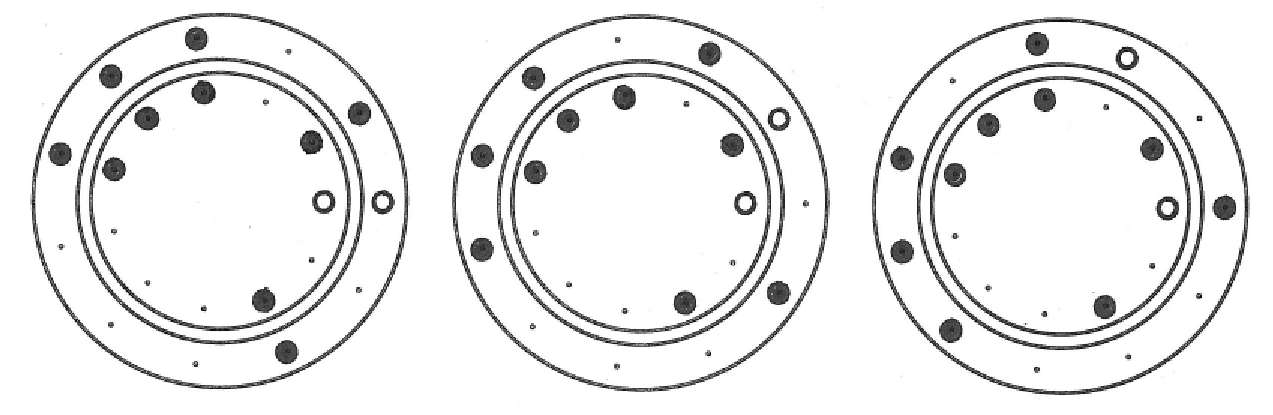,width=80mm}}}
\caption{\small{Following Perron, we illustrate our use of his
    theorems for $d = 11$. Reste are marked with bullets or in the
    exceptional case $0$ with a ring, while Nichreste are marked with
    small dots. Shifting the outer rim by an integer we find that the
    Reste line up in $(d+1)/4$ cases, while Nichtreste line up in
    $(d-3)/4$ cases. However, for our argument it is important that
    $0$ is treated separately. It lines up with a Rest if the shift is
    a quadratic residue (like $1$), and with a Nichtrest if the shift
    is a non-residue (like $2$).}}
\label{fig:Perron}
\end{figure}

Figure \ref{fig:Perron} may be helpful to follow the argument in the
subsequent proof of Lemma~\ref{lemma1}.

\begin{spiproof} 
First we calculate 
\begin{alignat}{5}
  \bra{{\bf v}} X^{-2j}\ket{{\bf v}}
  &{}= \sum_{k=0}^{d-1}v^*_kv_{k+2j}\nonumber\\
 &{}= v^*_0v_{2j} + v^*_{-2j}v_0 + \sum_{I}v^*_kv_{k+2j} 
+ \sum_{I\!I}v^*_kv_{k+2j} + \sum_{I\!I\!I}v^*_kv_{k+2j} \ , \label{summor} 
\end{alignat}
where $\sum_I$ runs over all $k$ such that 
\begin{equation}
  \textstyle  \legendre{k}{d}=\legendre{k+2j}{d},\label{eq:fall1}
\end{equation}
$\sum_{I\!I}$ runs over all $k$ such that 
\begin{equation}
  \textstyle  \legendre{k}{d}= 1
  \quad\text{and}\quad
  \legendre{k+2j}{d} = - 1, \label{eq:fall2}
\end{equation}
and $\sum_{I\!I\!I}$ runs over all $k$ such that 
\begin{equation}
  \textstyle\legendre{k}{d} = - 1
  \quad\text{and}\quad
  \legendre{k+2j}{d} = 1 . \label{eq:fall3} \end{equation}
Using the properties of the vector it follows that 
\begin{equation}
  \bra{{\bf v}} X^{-2j}\ket{{\bf v}}
   = - \sqrt{x_0}v_{2j} + \sqrt{x_0}v^*_{-2j} + \sum_{I} 1 - \sum_{I\!I}1/x_1^2 - \sum_{I\!I\!I}x_1^2. 
\end{equation}
We only need to know the number of terms in the sums. To proceed we will need the fact that 
\begin{equation}
  \legendre{2}{d} =
  \begin{cases}
     -1, &\text{if $d= p\equiv 3 \bmod 8$;}\\
     1, &\text{if $d = p\equiv 7 \bmod 8$.}
  \end{cases}\label{eq:Leg2} 
\end{equation}
This is where the proof splits into four cases. We give the argument for the case when 
$d\equiv 3\bmod 8$ and $j$ is a 
quadratic residue. The first two terms in \eqref{summor} are easily evaluated. We use 
\eqref{Leg1}, \eqref{eq:Leg2}, and the assumption that $j$ is a quadratic residue 
to compute 
\begin{equation}
  \legendre{-2j}{d}  =  -\legendre{2j}{d}  = \legendre{j}{d} = 1.
\end{equation}
From the way the vectors are defined it then follows that  
\begin{equation}
  - \sqrt{x_0}v_{2j} + \sqrt{x_0}v^*_{-2j}
    = - 2 \sqrt{x_0}v_{2j} =  2\sqrt{x_0}/x_1.
\end{equation}
For the remaining three sums we have to know how often the cases \eqref{eq:fall1}, 
\eqref{eq:fall2}, and \eqref{eq:fall3} occur, and for this we need Perron's S\"atze. 
Consider first $\sum_I$. There are $(d+1)/4$ pairs of Reste connected by the given 
shift, and also $(d-3)/4$ pairs of Nichtreste. But because $2j$ is a Nichtrest its 
negative is a Rest. Therefore the pair $(-2j, 0)$ has been included in the counting, 
so we have to subtract one from the sum; see Figure \ref{fig:Perron}. 
Hence 
\begin{equation}
  \sum_I 1 = \frac{d+1}{4} + \frac{d-3}{4} - 1 = \frac{d-3}{2}.
\end{equation}
For $\sum_{I\!I}$ we note that there are $(d+1)/4$ pairs of 
one Rest and one Nichrest, but since $2j$ is a Nichtrest this includes the pair 
$(0,2j)$, and again we have to subtract one from the sum. Hence 
\begin{equation}
  \sum_{I\!I}1/x_1^2 = \frac{d-3}{4}\frac{1}{x_1^2}.
\end{equation} 
For $\sum_{I\!I\!I}$ we are counting Nichtrest-Rest pairs and no 
complications occur. So 
\begin{equation}
  \sum_{I\!I\!I}x_1^2 = \frac{d+1}{4}x_1^2 .
\end{equation} 
Equation \eqref{fall1} follow. The proof of the remaining three cases is similar. 
\end{spiproof}

We state the main result of this section as a proposition: 

\begin{proposition} 
Legendre vectors obeying the $X$-overlap equation exist for
any $d = p\equiv 3\bmod 4$.
\end{proposition}

\begin{spiproof}
For $\legendre{j}{d}= 1$ the $X$-overlap equation is
\begin{equation}
  \bra{{\bf v}}X^{-2j}\ket{{\bf v}} =
  (\sqrt{d+1}+1)x_1^2.\label{eq:Xigen}
\end{equation} 
For $d\equiv 3\bmod 8$ we set this equal to the right hand side of \eqref{fall1} and obtain an equation with 
the solution  
\begin{equation}
  x_1 = \frac{\sqrt{-(\sqrt{d+1} + 1)} - 1}{\sqrt{x_0}}. \label{eq:x3}
\end{equation}
For $d = p\equiv 7\bmod 8$ we similarly find the solution 
\begin{equation}
  x_1 = \frac{\sqrt{-(d-3)(\sqrt{d+1} + 1)}-x_0}{\sqrt{d+1}\sqrt{x_0}}.\label{eq:x7}
\end{equation}
Although it is not obvious by inspection, a calculation verifies that these numbers are indeed phase factors.
\end{spiproof}

Hence we have found
Legendre vectors that solve the $X$-overlap equation for any $d =
p\equiv  3\mod 4$. But they are SIC vectors only if $d = 3$, $d = 7$,
or $d = 19$. We remark that if $d > 3$ the absolute degree of the number field
$\Q(\sqrt{x_0}x_1)$ equals $4$.

\section{Conclusions}

The $X$-overlap equation (\ref{eq:Xoverlap}) for the phase factors
entering a vector obeying our Ansatz
\eqref{eq:Ansatz1}--\eqref{Ansatz2}, or its counterpart for even
dimensions given in \cite{BGM}, are identities obeyed by certain Stark
units and their square roots in the (many) cases where this has been
tested. Our question was whether these identities determine the phase
factors to be the very square roots of Stark units that enter a SIC
vector of this form? The answer is in the negative. At least in prime
dimensions of the form $d = p\equiv 3\mod 4$ there are spurious
solutions to the equations that are Legendre vectors in the sense of
Khatirinejad, but most likely not SIC vectors unless $p = 7$, $19$.

It should be said, however, that our Gr\"obner basis calculations
showed that there are other dimensions where no spurious solutions
occur.  Indeed, the only spurious solutions we came across were the 
Legendre vectors. We conclude that the $X$-overlap equation merits 
further study.

\bmhead{Acknowledgements}
We thank the organizers of the Hadamard 2025 conference in Sevilla,
which inspired this work. We also thank Gary McConnell and Marcus 
Appleby for discussions and useful comments.

\bmhead{Funding}
  Ingemar Bengtsson acknowledges support by the Digital Horizon Europe
  project FoQaCiA, Foundations of quantum computational advantage, GA
  No. 101070558, funded by the European Union, NSERC (Canada), and
  UKRI (UK).

  The work of Markus Grassl is carried out under the `International
  Centre for Theory of Quantum Technologies 2.0: R\&D Industrial and
  Experimental Phase’ project (contract
  no.~FENG.02.01-IP.05-0006/23). The project is implemented as part of
  the International Research Agendas Programme of the Foundation for
  Polish Science, co-financed by the European Funds for a Smart
  Economy 2021-2027 (FENG), Priority FENG.02 Innovation-friendly
  environment, Measure FENG.02.01.

\hypersetup{breaklinks=true}

\end{document}